\documentclass[pra,print,superscriptaddress,twocolumn]{revtex4}
\usepackage{amssymb}
\usepackage{amsmath}
\usepackage{epsfig}
\usepackage{color}
\usepackage{graphics, graphicx}
\usepackage{bbold}
\usepackage{psfrag}
\usepackage{mathcomp}
\usepackage{subfigure}
\usepackage{verbatim}
\usepackage{color}
\usepackage[colorlinks,citecolor=blue]{hyperref}

\begin{document}

\title{Magnetic order and strongly-correlated effects in the one-dimensional
Ising-Kondo lattice}
\author{Xiaofan Zhou}
\affiliation{State Key Laboratory of Quantum Optics and Quantum Optics Devices, Institute
of Laser Spectroscopy, Shanxi University, Taiyuan 030006, China}
\affiliation{Collaborative Innovation Center of Extreme Optics, Shanxi University,
Taiyuan 030006, China}
\author{Jingtao Fan}
\thanks{fanjt@sxu.edu.cn}
\affiliation{State Key Laboratory of Quantum Optics and Quantum Optics Devices, Institute
of Laser Spectroscopy, Shanxi University, Taiyuan 030006, China}
\affiliation{Collaborative Innovation Center of Extreme Optics, Shanxi University,
Taiyuan 030006, China}
\author{Suotang Jia}
\affiliation{State Key Laboratory of Quantum Optics and Quantum Optics Devices, Institute
of Laser Spectroscopy, Shanxi University, Taiyuan 030006, China}
\affiliation{Collaborative Innovation Center of Extreme Optics, Shanxi University,
Taiyuan 030006, China}

\begin{abstract}
We investigate the magnetic order and related strongly-correlated effects in
an one-dimensional Ising-Kondo lattice with transverse field. This model is
the anisotropic limit of the conventional isotropic Kondo lattice model, in
the sense that the itinerant electrons interact with the localized magnetic
moments via only longitudinal Kondo exchange. Adopting the numerical
density-matrix-renormalization group method, we map out the ground-state
phase diagram in various parameter spaces. Depending on the Kondo coupling
and filling number, three distinct phases, including a metallic
paramagnetic, a metallic ferromagnetic, and a gapped spin-density wave
phase, are obtained. The spin-density wave is characterized by an ordering
wave vector which coincides with the nesting wave vector of the Fermi
surface. This makes the corresponding magnetic transition a spin analog of 
the Peierls transition occurring in the one-dimensional metal. Moreover, by
analyzing the momentum distribution function and charge correlation
function, the conduction electrons are shown to behave like free spinless
fermions in the ferromagnetic phase. We finally discuss the effect of the
repulsive Hubbard interaction between conduction electrons. Our work
enriches the Kondo physics and deepens the current understanding of the
heavy fermion compounds.
\end{abstract}

\maketitle

\section{Introduction}

As one of the most important canonical models in condensed matter systems,
the Kondo lattice model (KLM) is used to describe the process where a
conduction band, occupied by itinerant fermions, interacting with a lattice
of localized magnetic moments \cite{KLM1}. The KLM is relevant to a wide
class of real materials called heavy fermion system \cite%
{KLM2,KLM3,superconducting,KLM4,KLM5,KLM6,KLM7,HF01,HF02,HF03}, in which huge quasiparticle mass can emerge
due to the formation of Kondo singlets \cite{HF1,HF2}. The study of such
heavy fermion materials is believed to be critical for understanding diverse
phenomena such as quantum criticality \cite{KLM4,Critical1,Critical01,Critical02,Critical03,Critical2,Critical04,Critical05,Critical06} and
high-$T_{c}$ superconducting compounds \cite{KLM2,KLM3,superconducting}. While the
well-solved single-impurity Kondo problem provides some useful concepts for
the KLM, there are aspects where the single-impurity results have no analog
in the lattice case \cite{SKondo1,SKondo2,SKondo3}. For example, except
for a few exact results in certain limits \cite%
{ExactKLM1,ExactKLM2,ExactKLM3}, the dynamical details of the interplay
between the local Kondo physics and the nonlocal
Ruderman-Kittel-Kasaya-Yosida (RKKY) interaction remains elusive. As a
consensus that has been reached, under zero temperature, the one-dimensional
(1D) KLM presents three distinct magnetic phases: the anti-ferromagnetic
phase (AFM) with fully opened spin and charge gaps, the metellic
paramagnetic phase (PM) with RKKY correlations, and the metellic
ferromagnetic phase (FM) \cite{KLM3,KLMP1,KLMP2,KLMP3,KLMP4,KLMP5,KLMP6}.
The AFM is stable down to zero Kondo coupling when the conduction band is
half-filled, whereas the PM-to-FM transition can occur only away from
half-filling \cite{HalfKLM1,HalfKLM2,HalfKLM3}.

The KLM respects a SU(2) rotational symmetry due to the isotropic Kondo
interaction in the spin space. However, by adding anisotropy between the
Kondo couplings, numerous interesting physics, which is beyond the isotropic
model, can emerge \cite{AKLM01,AKLM02,AKLM03}. For example, the KLM with
properly tuned anisotropy has been utilized to understand the physics of the
dissipative two-state systems \cite{AKLM1,AKLM2}. Moreover, it has been shown
that the ferromagnetic Kondo couplings with easy-plane anisotropy may yield
anomalous singlet formation \cite{AKLM3}. Among various kinds of anisotropy
introduced in the KLM, the Ising-Kondo lattice model (IKL) with only
longitudinal Kondo coupling takes on a special role \cite%
{IKL1,IKL2,IKL3,IKL4,IKL5,IKL6,IKL7}. Its importance is due both to its
simple Ising-type coupling form, amenable to both analytical and numerical
treatments \cite{IKL5}, and to its relevance to a series of real materials
\cite{IKL1,IKL2,IKL3,IKL4}. The IKL was first proposed to describe the
concurrence of large specific heat jump and weak antiferromagnetism in URu$_{2}$Si$_{2}$ \cite{IKL1}. Recently, some thermodynamic properties of the
IKL \cite{IKL5}, including the antiferromagnetic topological character \cite%
{IKL6} and the emergent strange metal behaviors \cite{IKL7}, were discussed
on two-dimensional lattices. However, a systematic understanding of the
magnetic order in the IKL under various conduction electron concentrations,
especially for the 1D case, is still lacking.

In this work, we investigate the ground-state properties of the 1D IKL with
transverse field by using the numerical density-matrix-renormalizationgroup
(DMRG) method \cite{dmrg1,dmrg2}. The competition between the nonlocal RKKY
mechanism and the local Kondo physics leads to the emergence of magnetic
correlations and, consequently, various magnetic orders. Apart from the
metallic PM and FM, a gapped spin-density-wave phase (SDW) with long-range
order is unraveled. The ordering wave vector characterizing the SDW is found
to be nothing but the nesting wave vector of the Fermi surface constructed
by the conduction electrons, making the SDW a Peierls-like insulator. We
also discuss the effect of repulsive Hubbard interaction imposed on the
magnetic orders. It is revealed that the Hubbard interaction influences the
magnetic phases in a way much like the Kondo coupling; it can trigger both
the PM-to-SDW and SDW-to-FM transitions, depending on the filling
number of the conduction electrons. Remarkably, we find that, deep inside
the FM region, the conduction electrons behave like free spinless fermions,
characterized by a distinct redistribution of the electron quasi-momentum
and the corresponding signals in the charge correlation function.

\section{Model and method}

\label{sec:system}

The IKL in consideration can be described by the
following Hamiltonian ($\hbar =1$ throughout)%
\begin{eqnarray}
\hat{H} &=&-t\sum_{\left\langle i,j\right\rangle ,\sigma }\hat{c}_{i,\sigma }^{\dag
}\hat{c}_{j,\sigma }+h\sum_{j}\hat{s}_{j}^{z}\hat{S}_{j}^{z}+\Delta \sum_{j}\hat{S}_{j}^{x}  \notag
\\
&&+\frac{U}{2}\sum_{j}\hat{n}_{j,\uparrow }\hat{n}_{j,\downarrow }  \label{H1}
\end{eqnarray}%
where $\hat{c}_{j,\sigma }^{\dag }$ ($\hat{c}_{j,\sigma }$) is the creation
(annihilation) field operator of the conduction electron with spin $\sigma $
(=$\uparrow ,\downarrow $) at lattice site $j$. The conduction electrons can
hop between adjacent sites $\left\langle i,j\right\rangle $ with the hopping
rate $t$. $\hat{S}_{j}^{z}$ denotes the $z$-component of the spin-1/2 operator for the localized magnetic moment. The spin operator $%
\hat{s}_{j}^{z}$\ for the conduction electrons is defined by $\hat{s}_{j}^{z}=\sum_{\tau
,\acute{\tau}}\hat{c}_{j,\tau }^{\dag }\sigma _{\tau ,\acute{\tau}}^{z}\hat{c}_{j,\acute{\tau}}/2$ where $\sigma ^{z}$ is the Pauli-$z$ matrix. With these definitions,
the second term of Hamiltonian (\ref{H1}) denotes the longitudinal Kondo
interaction between the conduction electron and the localized moment at the
same site $j$. The interaction strength $h$ is assumed to be positive,
implying an antiferromagnetic Kondo coupling. Note that for the IKL
considered here, the ferromagnetic coupling with $h<0$ is
connected to the antiferromagnetic case through a simple spin rotation, which is in contrast to the conventional isotropic KLM. The latter can exhibit dramatic different Kondo physics
depending on the sign of the coupling strength \cite{KLM8,KLM9}. A
transverse field $\Delta $, introducing additional nonadiabaticity, is applied on the localized moments \cite{IKL1}. The direct repulsive interaction between
conduction electrons is also included through the Hubbard-$U$ term. In this
work, our main focus will be on the ground-state properties of the
conduction electrons. The local moments construct the same magnetism with
conduction electrons due to the longitudinal Kondo coupling term $\sim
\hat{s}_{j}^{z}\hat{S}_{j}^{z}$. In the following discussion, we set the energy scale by
taking $t=1$, and we also take $\Delta =1$ unless otherwise specified.

The Hamiltonian (\ref{H1}), with vanishing transverse Kondo couplings (i.e., $\hat{s}_{j}^{-}\hat{S}_{j}^{+}+$\ H.c.$\rightarrow 0$), represents essentially the
anisotropic limit of the conventional KLM  \cite{KLM1}. In the adiabatic limit with $\Delta =0$, the IKL is exactly
sovable since the local moment at each site is conservative obeying $%
[\hat{S}_{j}^{z},\hat{H}]=0$. The thermodynamic properties of the IKL under this
condition has been preliminarily touched by Quantum Monte Carlo simulations
\cite{IKL5}. A nonzero transverse field $\Delta $, however, breaks the
conservativeness of $\hat{S}_{j}^{z}$ and effectively adds quantum fluctuations to
the localized moments. In this sense, the localized moments in IKL can be
alternatively viewed as lattice degrees of freedom which impose dynamic
potentials on the conduction electrons. The interactions between fermions
and lattice degrees of freedom become specially important in 1D due to the
perfect Fermi surface nesting \cite{nest1,nest2}. One of the most noted
consequences, for example, is the Peierls transition, characterized by the
formation of a charge-density wave ordered at the nesting wave vector \cite%
{Peierls}. In light of the Ising-type interaction between conduction
electrons and localized moments, in the IKL, we expect that an analogous SDW
instability of the conduction electrons can be reached through similar
mechanisms. This point will be further elucidated in the subsequent sections.

The magnetic order of the conduction electrons can be effectively
characterized by the spin structure factor,%
\begin{equation}
S(k)=\frac{1}{L}\sum_{l,j}\left\langle \hat{s}_{l}^{z}\hat{s}_{j}^{z}\right\rangle
e^{i(l-j)k}
\end{equation}
where $L$ is the number of lattice sites and $\left\langle ...\right\rangle $
denotes the ground-state average. The peak position in $S(k)$, denoted by an
ordering wave vector $k^{\max }$, characterizes the spatial variation of
spin orientations projected into the $z$ direction. Alternatively, we can
characterize the magnetic order using the spin correlation function in real
space
\begin{equation}
s(r)=\frac{1}{L}\sum_{l}\left\langle \hat{s}_{l}^{z}\hat{s}_{l+r}^{z}\right\rangle
\end{equation}
where $r$ is the distance between different sites. The correlation function $s(r)$ has the advantage of intuitively showing the spatial distribution of
the spin correlations. The equivalence between $S(k)$ and $s(r)$ can be
clearly found by noting that they constitute a Fourier transform pair. In
the spirit of Landau's paradigm, the magnetic ordering is characterized by
spontaneously broken symmetry at some ordering wave vector $k^{\max }$. This
implies the existence of long-range order, which can be identified by
scaling the spin structure factor \cite{Long}:%
\begin{equation}
\lim_{L\rightarrow \infty }\frac{1}{L}S(k^{\max })>0.
\label{SkL}
\end{equation}
The long-range nature is also manifested in the correlation function $s(r)$
by $\lim_{r\rightarrow \infty }s(r)\neq 0$. Therefore, we can identify
different magnetic phases according to Eq. (\ref{SkL}) by extrapolating the
spin structure factor to the thermodynamic limit $L\rightarrow \infty $. In
this way, we may obtain the $k^{\max }$-ordered SDW for $k^{\max }\neq 0$
and the FM for $k^{\max }=0$. If $S(k)/L$ is extrapolated to zero in the
thermodynamic limit, the corresponding phase loses any magnetic long-range
order and is thus termed PM. We emphasize that, due to the space-inversion
symmetry of the system, any peaks in $S(k)$, if exist, should be exactly
symmetric about $k=0$.

Here, we perform state-of-the-art DMRG calculations to compute the many-body
ground state of the system, with which various physical observable can be
obtained. In our numerical simulations, the filling number $\rho =N/L$ is a
good quantum number which can be varied from zero to one. Here $N$ is the
total number of conduction electrons. We set lattice size up to $L=60$, and
work with open boundary conditions.\ For each lattice size, we retain 600
truncated states per DMRG block and perform 20 sweeps with a maximum
truncation error of $\sim 10^{-9}$.

\begin{figure}[tp]
\includegraphics[width=7.5cm]{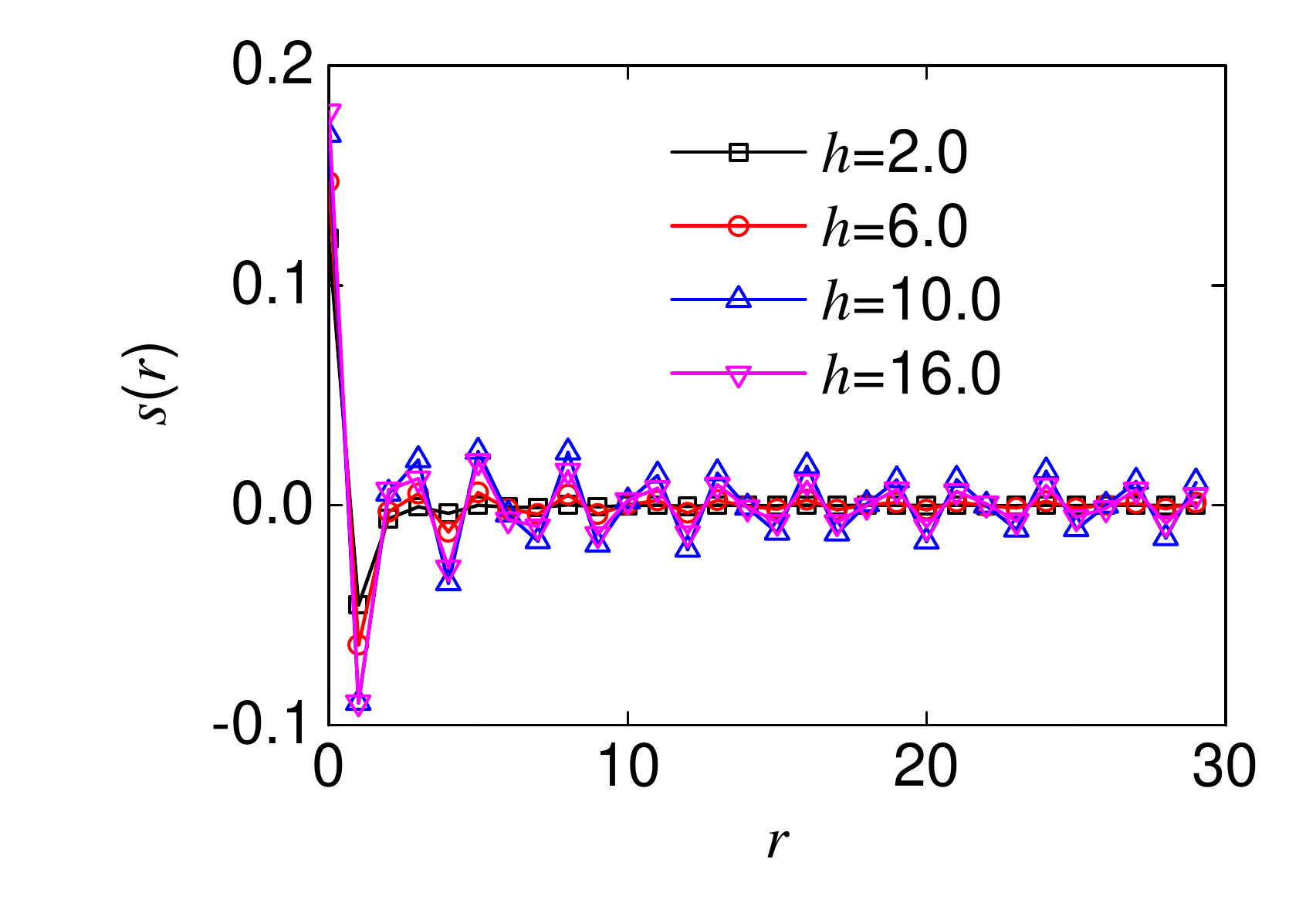}
\caption{The correlation function $s(r)$ for systems with $\protect\rho %
=0.75 $, $L=60$, and a varying $h$.}
\label{Fig1}
\end{figure}

\section{Magnetic correlations induced by Kondo coupling}
\label{sec:results1}

We first clarify the effect of the Kondo coupling $h$ by setting $%
U=0$. Before showing the numerical results, some qualitative insights can be
gained by inspecting different parameter regimes. For the coupling strength
comparable with or less than the Fermi energy of the conduction electrons, the dominant ordering force turns out to be the RKKY
correlation with Ising anisotropy, which compete with the quantum
fluctuations induced by the transverse field $\Delta $. Increasing $h$
further to the intermediate coupling regime, the disorder created by $\Delta
$ is irrelevant and the interplay between the RKKY-type interactions and the
local Kondo physics becomes important. At extremely strong-coupling $h\gg 1$, the conduction electrons and the local moments are tightly bound together to form local spin singlets, giving rise to distinct local Kondo physics. We thus expect rich long-range magnetic
orders may emerge throughout the whole range of the coupling strength.

\begin{figure}[tp]
\includegraphics[width=9cm]{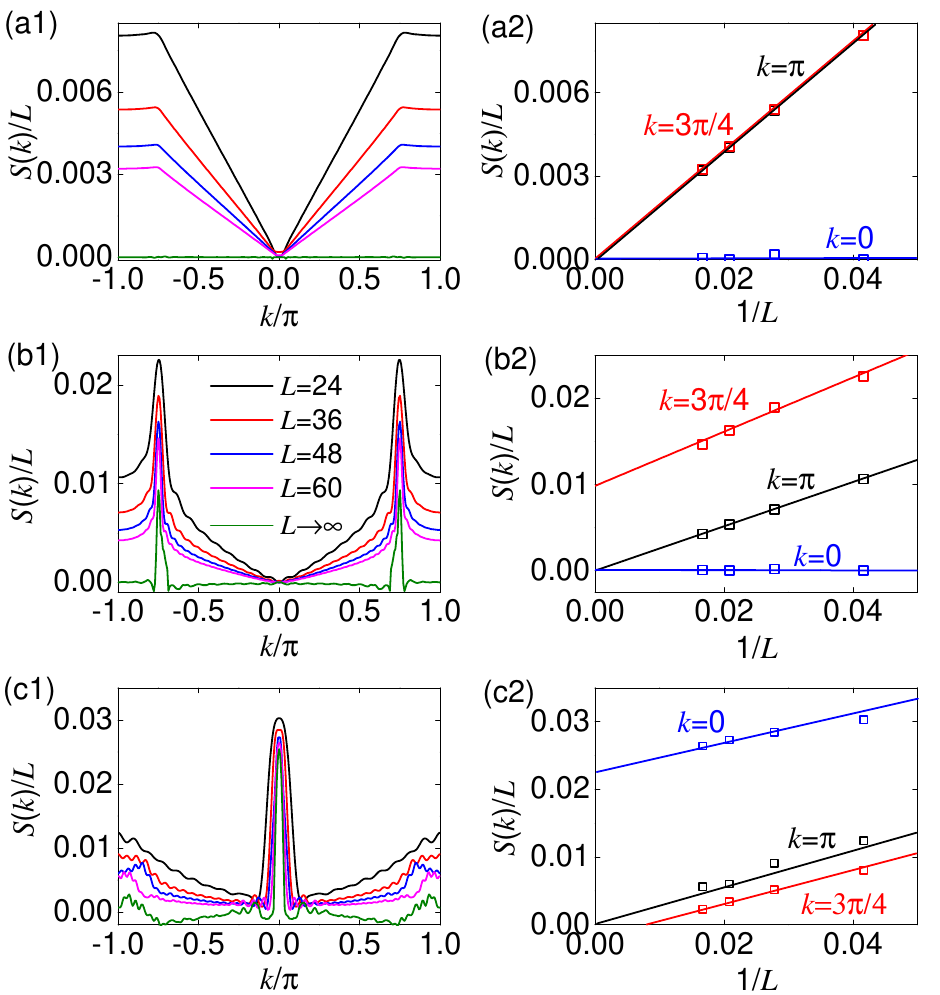}
\caption{(a1)-(c1) The scaled spin structure factor $S(k)/L$ and (a2)-(c2)
the corresponding finite-size scalings at three characteristic wave vectors
for systems with (a) $h=2.0$, (b) $h=10.0$, and (c)\ $h=22.0$. Different system
sizes are characterized by lines with different colors. In these figures, we
set $\protect\rho =0.75$ and $U=0$. }
\label{Fig2}
\end{figure}

\subsection{Phase transitions and long-range order}

Let us detailedly discuss the magnetic phase behaviors of the IKL in different coupling regimes. Figure~\ref{Fig1} shows the correlation function $s(r)$ with $\rho =0.75$ and different $h$. When $h$ is
relatively weak, the spin correlation rapidly decays to zero as the distance $r$ increases. For larger $h$, we
observe a persistent oscillation of $s(r)$ instead. This non-decaying behavior of
spin correlation signals the emergence of a long-range magnetic order.

The transitions between different magnetic phases becomes clearer if we investigate the corresponding spin structure factor $S(k)$. The left column of Fig.~\ref{Fig2} [ Figs.~\ref{Fig2}(a1)-\ref{Fig2}(c1)] plot $S(k)/L$ for three different values of $h$ with $\rho =0.75$.
The results for different system sizes are labeled by lines with different
colors. The values of $L\rightarrow \infty $ are obtained by the
standard finite-size scaling. The right column of Fig.~\ref{Fig2} [Figs.~\ref{Fig2}(a2)-~\ref{Fig2}(c2)] demonstrate the scaling details at three characteristic wave vectors. We first focus on the weak
coupling strength with $h=2.0$ [Fig.~\ref{Fig2}(a1)]. It is shown that $S(k)/L$
develops weak peaks at $k=k^{\max }=\pm 2k_{F}$ for any finite system sizes, resulting from a RKKY-type correlation. Here $k_{F}=\pi \rho /2$ is the
Fermi momentum of the conduction electron. As the system size increases,
however, the peaks become weaker and weaker and eventually disappear in the
thermodynamic limit $L\rightarrow \infty $. This implies that the magnetic correlations built up in this regime is essentially short
range, and consequently give rise to the appearance of PM. Something interesting happens if the coupling strength increases. The spin structure factor with $h=10.0$ is shown in Fig.~\ref{Fig2}(b1). It is found that, while the structure with $k^{\max }=\pm 2k_{F}$
still exists, the peaks become much higher and sharper compared to $h=2.0$. Accordingly, the scaled structure factor $S(k=k^{\max })/L$ is extrapolated to a finite value in the thermodynamic limit [see the red line
in Fig.~\ref{Fig2}(b2)], indicating the existence of a SDW which is ordered
at $k=k^{\max }$. This is in sharp contrast with the isotropic KLM. In that case, the conduction electrons lose
the true SDW order due to additional fluctuations, although short-range magnetic correlations still exist \cite{KLM3}. Increasing $h$ further, the conduction
electrons and localized moments start to be bound together and the tendency of
local Kondo singlet formation overwhelms that of the RKKY-type correlation. As a
result, a sharp peak at $k=0$ grows up whereas density waves at other wave
vectors are greatly suppressed, as shown in Fig.~\ref{Fig2}(c1). This
structure of $S(k)$ unambiguously demonstrates the ferromagnetism. We thus find
that as $h$ increases, the system starts from PM, and subsequently traverses
the SDW, and eventually evolves to the FM.

\begin{figure}[tp]
\includegraphics[width=9cm]{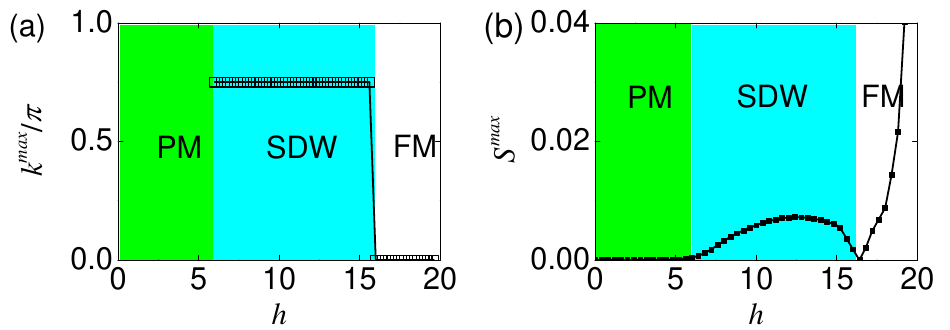}
\caption{(a) The ordering wave vector $k^{\max }$ and (b) the ordering
strength $S^{\max }$ as functions of $h$ with $\protect\rho =0.75$ and $U=0$%
.\ }
\label{Fig3}
\end{figure}

To provide a more precise picture, we monitor the ordering wave vector $k^{\max }$ as a function of the coupling strength for $\rho =0.75$. Considering the inherent inversion
symmetry of $S(k)$ about $k=0$, we here focus on the $k^{\max }\geqslant 0$
part. As shown in Fig.~\ref{Fig3}(a), $k^{\max }$ is not well defined for $%
h<5.8$ due to the general vanishing of $S(k)/L$ in the limit of $%
L\rightarrow \infty $, which is the character of the PM. Increasing $h$ above $5.8$,
a well-defined $k^{\max }$ emerges at $2k_{F}$, and keeps at that value
until it drops to zero where the FM is formed. The feature that $%
k^{\max }$ keeps constant before reaching FM is different
from the isotropic KLM, where the peaks in $S(k)$ can move continuously toward zero as $h$ increases \cite{KLM9}.
Figure~\ref{Fig3}(b) plots $S^{\max}$, which we dub ordering strength, in terms of  $h$. $S^{\max }$ is the maximum of the scaled spin structure factor $\lim_{L\rightarrow \infty}S(k)/L$ within the Brillion zone. With this definition, we find the relation $S^{\max}=\lim_{L\rightarrow \infty}S(k^{\max })/L$ applies in the ordered phase. It follows that $S^{\max }$,
together with the value of $k^{\max }$, serve as complete order parameters characterizing the ordering process of different magnetic phases. As shown in Fig.~\ref{Fig3}(b), while $S^{\max }$ remains zero in
the PM, it gradually increases from zero, reaching the maximum, and
then decreases until vanishes at the critical point $h_{c}=16.4$. Above $h_{c}$, the system enters the FM with monotonically increased $S^{\max }$. 

\begin{figure}[tp]
\includegraphics[width=9cm]{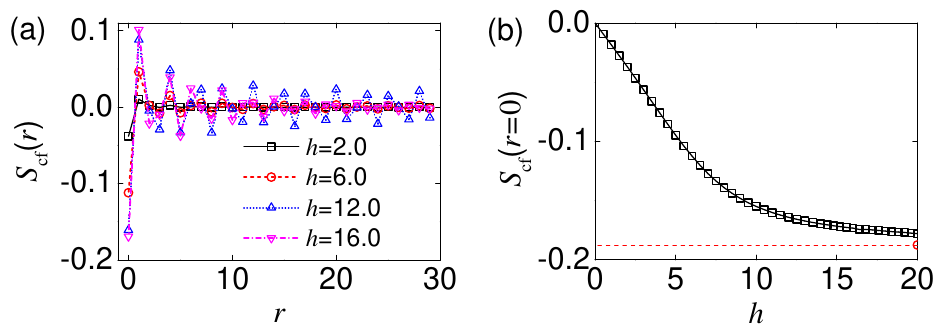}
\caption{(a) The correlation function $S_{\text{cf}}(r)$ for different $h$.\
(b) The on-site value $S_{\text{cf}}(0)\ $as a function $h$. The other
parameters in these figures are $\protect\rho =0.75$, $U=0$, and $L=60.$ The red dashed line in (b) denotes the value of the perfect on-site singlets.}
\label{Fig4}
\end{figure}

\begin{figure}[b]
\includegraphics[width=9cm]{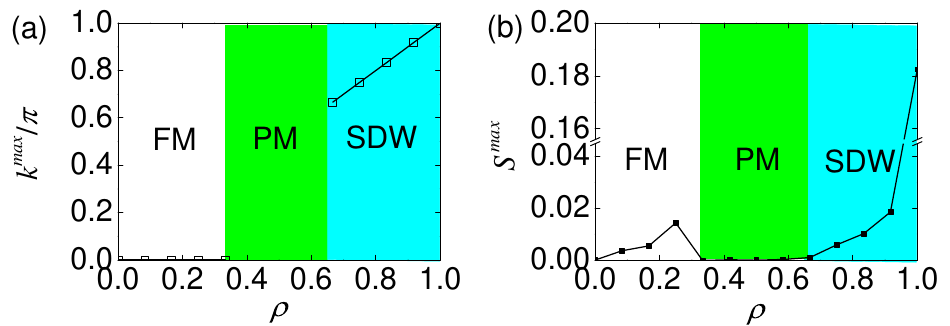}
\caption{(a) The ordering wave vector $k^{\max }$ and (b) the ordering
strength $S^{\max }$ as functions of $\protect\rho $ with $h=12.0$ and $U=0$.\ }
\label{Fig5}
\end{figure}

The non-monotonically behavior of $S^{\max }$ results from the competition between the
RKKY-type interaction and the Kondo singlet formation. The details of this competition is partially manifested in the correlation function between the conduction
electron and local moment:
\begin{equation}
S_{\text{cf}}(r)=\frac{1}{L}\sum_{l}\left\langle
\hat{s}_{l}^{z}S_{l+r}^{z}\right\rangle . 
\label{Scf}
\end{equation}
As defined in Eq.~\ref{Scf}, the correlation function $S_{\text{cf}}(r)$ measures the non-local character of the Kondo singlet formation \cite{KLMP2}. Figure~\ref%
{Fig4}(a) plots $S_{\text{cf}}(r)$ with a varying values of $h$. 
At weak coupling $h=2$, the hybridization between the conduction electron and local moment is ineffective, yielding a rather weak signal in $S_{\text{cf}}(r)$. For larger $h$, $S_{\text{cf}}(r)$ increases in height and it decays very slowly as $r$ increases, reflecting a sizeable extension of the Kondo singlet.
This long-range behavior of $S_{\text{cf}}(r)$ signals the appearance of a regime where the RKKY mechanism may work. Increasing the coupling strength further such that $h\gtrsim 12$, $S_{\text{cf}}(r)$ becomes more and more localized, implying that the size of the singlet reduces. Consequently, the RKKY mechanism starts to be suppressed. This point is further strengthen by examining the on-site
singlet correlation $S_{\text{cf}}(0)$ versus $h$. As shown in Fig.~\ref%
{Fig4}(b), as $h$ increases, $S_{\text{cf}}(0)$ approaches asymptotically
that of perfect on-site singlets $-3\rho /4$ (see the red dashed line in Fig.~\ref{Fig4}(b)).

\begin{figure}[tp]
\includegraphics[width=7.5cm]{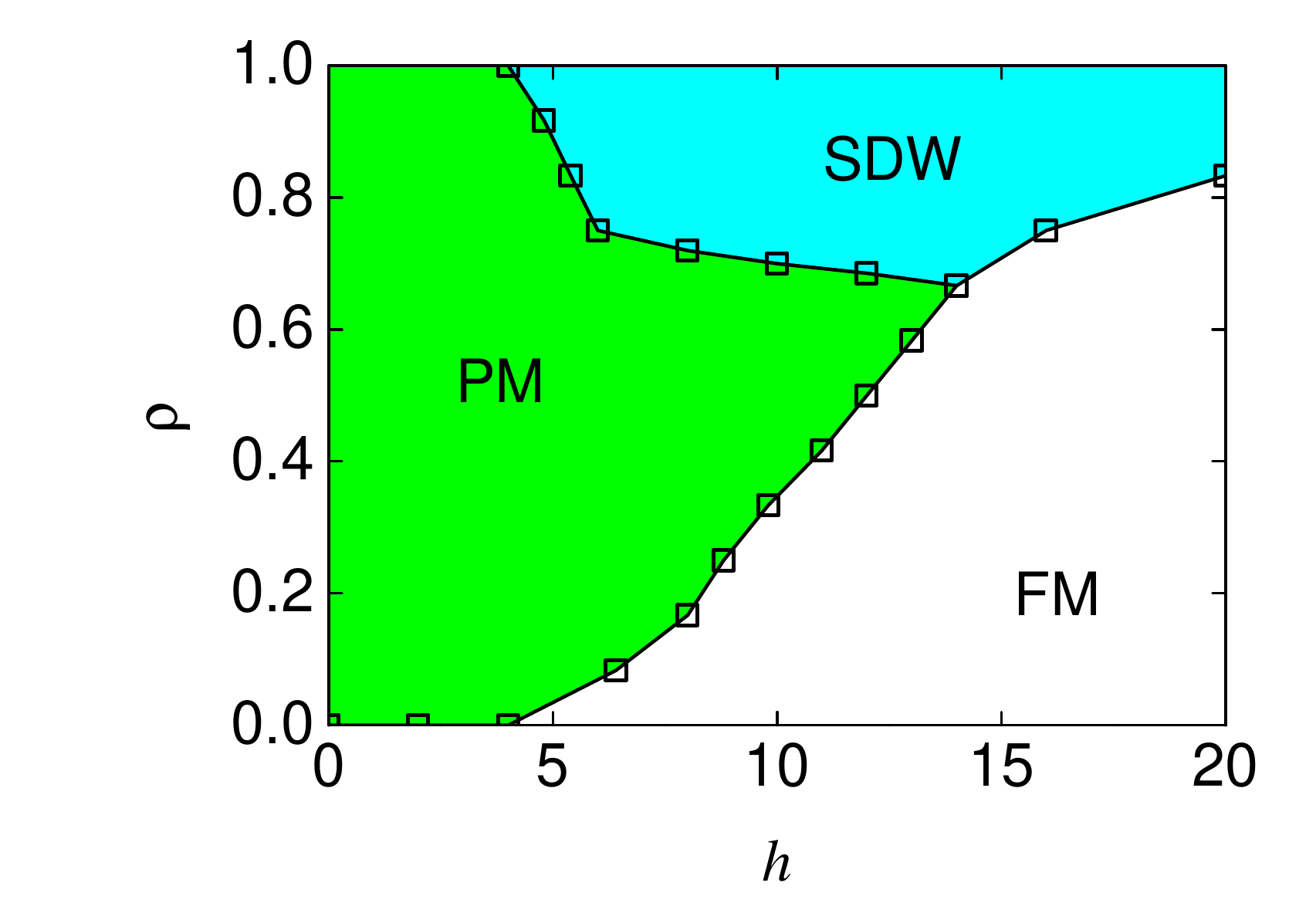}
\caption{The phase diagram in the $h-\protect\rho $ plane with $U=0$. The phase boundaries have been extrapolated to the thermodynamic limit $L\rightarrow \infty $.}
\label{Fig6}
\end{figure}

Up to now, our discussion is limited to a specific band filling of the
conduction electrons, namely $\rho =0.75$. We have observed that the
ordering wave vector of the SDW in this filling is fixed by the Fermi
momentum of the conduction electrons via $k^{\max }=\pm 2k_{F}$. In fact,
this relation applies to any fillings provided the system is within the SDW. To
demonstrate this, we plot $k^{\max }$ and $S^{\max }$ as functions of $\rho $
with $h=12.0$ in Figs.~\ref{Fig5}(a) and \ref{Fig5}(b), respectively. As an
immediate finding, while $k^{\max }$ keeps filling independent in the FM\
and PM, it follows $\rho $ linearly through $k^{\max }=\pm 2k_{F}$ in the
SDW region [see Fig \ref{Fig5}(a)]. Notice that the ordering wave vector $%
2k_{F}$ is exactly the nesting wave vector of the Fermi surface, which
characterizes the charge density wave in an Peierls insulator \cite{Peierls}%
. It is thus inferred that the appearance of the SDW in the 1D IKL can be
traced back to the perfect Fermi surface nesting effect. The latter drastically modifies
the band structure of itinerant electrons. From this point of view, the
formation of SDW in the IKL can be regarded as a magnetic analog of the
Peierls transition which occurs in the charge degree of freedom.

\subsection{Phase diagram}

With the understanding above, we map out the phase diagram in the $h-\rho $
plane in Fig.~\ref{Fig6}. As can be seen, the PM occupies most portion of
the phase diagram for weak coupling strength, due to the dominant quantum
fluctuations of the local moments. As $h$ increases, the ordered phases,
including FM and SDW with various ordering wave vectors, emerge. Whereas the
SDW is more favored when the filling number is closed to $\rho =1$, the FM
is formed away from this half-filling case. The ordering process for $h=12.0$
is illustrated in Fig.~\ref{Fig5}(b), where we find $S^{\max }$ increases
rapidly as $\rho $ approaches one. We emphasize that the SDW for $\rho =1$
is an antiferromagnetic insulator characterized by a N\'{e}el configuration
with finite spin and charge gaps \cite{IKL5}. Increasing $h$ above $14.0$,
the PM generally disappears and the SDW region starts to shrink. The
antiferromagnetic insulator for $\rho =1$, however, remains stable against
the ferromagnetization even in the $h \rightarrow \infty $ limit \cite{IKL5}.

\begin{figure}[tp]
\includegraphics[width=9cm]{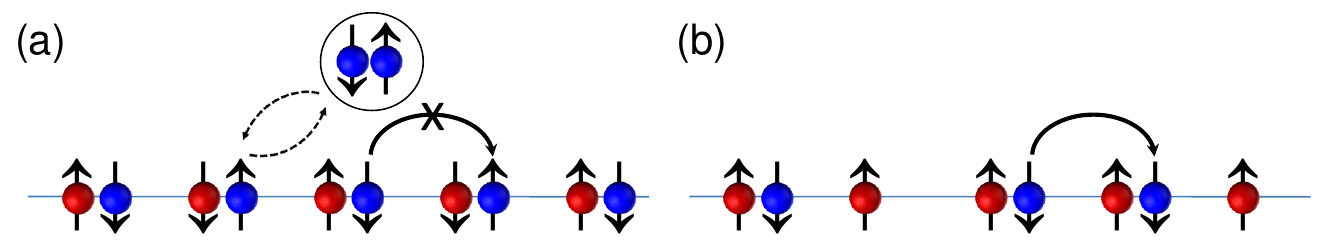}
\caption{Schematic illustration of the magnetic structures when the
conduction electrons are (a) around and (b) far away from half filling. In
these figures, the red and blue balls denote the localized moments and
conduction electrons, respectively. }
\label{Fig7}
\end{figure}

The mechanism of different magnetic structures closed to and far away from
half filling can be understood as follows. As illustrated in Fig.~\ref{Fig7}%
(a), in the strong coupling regime and around half filling, each neighbor
sites of the conduction electron is likely to be occupied by another one,
since double occupancy of the conduction electrons on the same site would
elevate the energy by $\Delta E\sim h/4$. As a result, the hopping between
neighbor sites is forbidden by the Pauli exclusion principle if the spins of
conduction electrons are aligned. However, if the spins form an
antiferromagnetic configuration, a virtual tunnelling process can occur which
lowers the kinetic energy of the conduction electrons by $\Delta E\sim
4t^{2}/h$, resembling closely the superexchange process in the Hubbard model
\cite{Long}. Far away from half filling, on the other hand, the conduction
electrons can freely move along the lattice without costing any additional
energy once their spins are aligned parallel to each other [see Fig.~\ref%
{Fig7}(b)]. 

\begin{figure}[b]
\includegraphics[width=7.0cm]{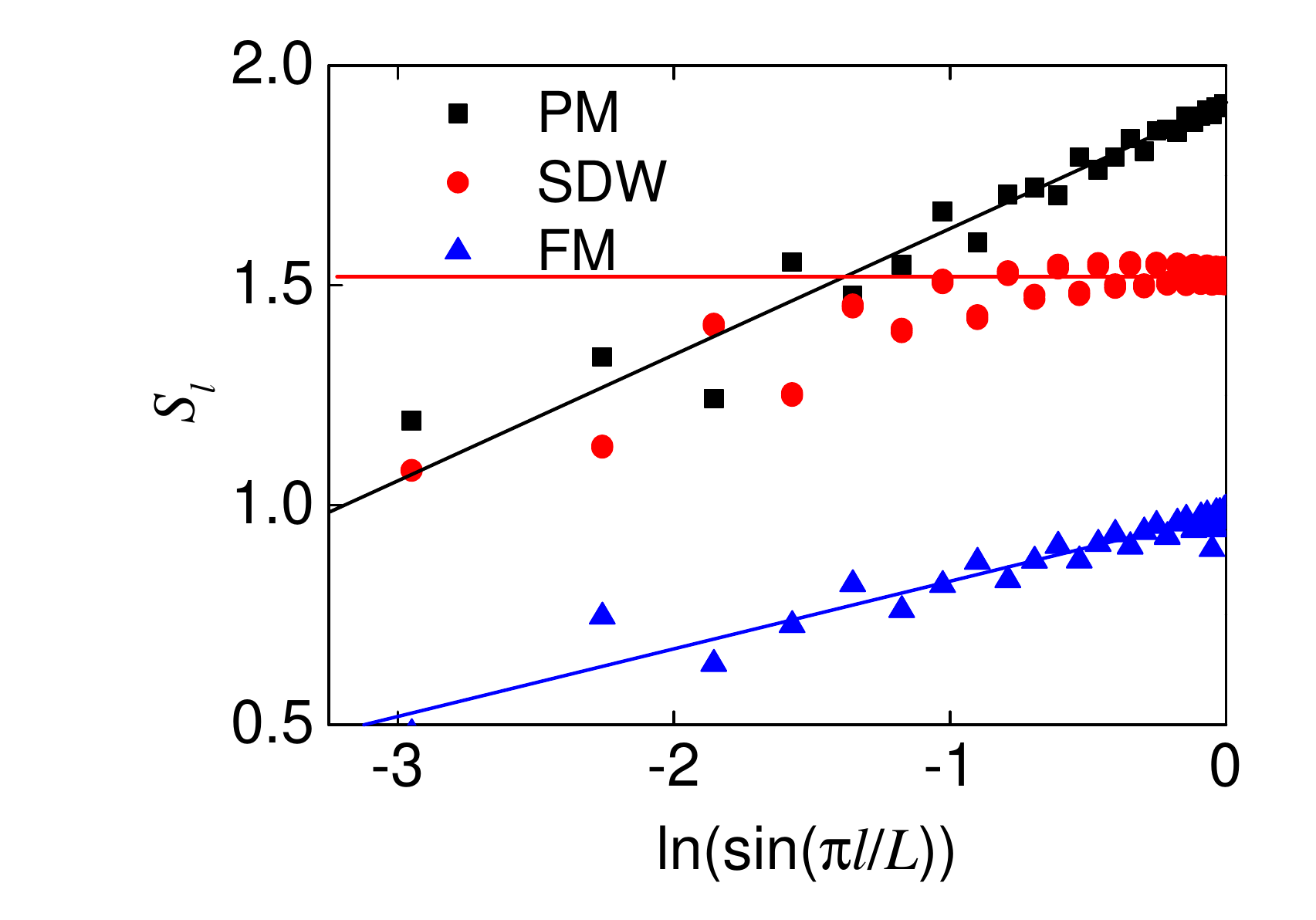}
\caption{The entanglement entropy $S_{l}$ in terms of the block size $l$,
for representative points of PM ($h=2.0$, $\protect\rho =0.75$), SDW ($h=12.0$, $%
\protect\rho =0.5$), and FM ($h=18.0$, $\protect\rho =0.5$). The scaling
behavior for different phases are plotted by lines with different colors. }
\label{Fig8}
\end{figure}

The above picture hints that the system is a gapped insulator
throughout the SDW (not only the half-filling case) and becomes metallic
when moving into FM and PM. This conjecture is verified by calculating the
scaling of the entanglement entropy between a block of size $l$ and the rest
of the system, $S_{l}=-$Tr$(\rho _{l}\ln \rho _{l})$, where $\rho _{l}=$Tr$%
_{L-l}\left\vert \Psi \right\rangle \left\langle \Psi \right\vert $ is the
reduced density matrix corresponding to the block. For 1D systems, gapped
phases follows an area law, and the entanglement entropy $S_{l}$ would\
saturate as $l$ increases\cite{entropy1}. $S_{l}$ in gapless phases,
however, does not saturate but grows logarithmically when increasing $l$ \cite%
{entropy2}. The scaling of $S_{l}$ for three representative points in the
phase diagram are shown in Fig.~\ref{Fig8}. While the presence of a gap in
the SDW is indicated by the saturation of $S_{l}$, the FM and PM are
both gapless since $S_{l}$ logarithmically diverges with the block's size.

\begin{figure}[tp]
\includegraphics[width=9cm]{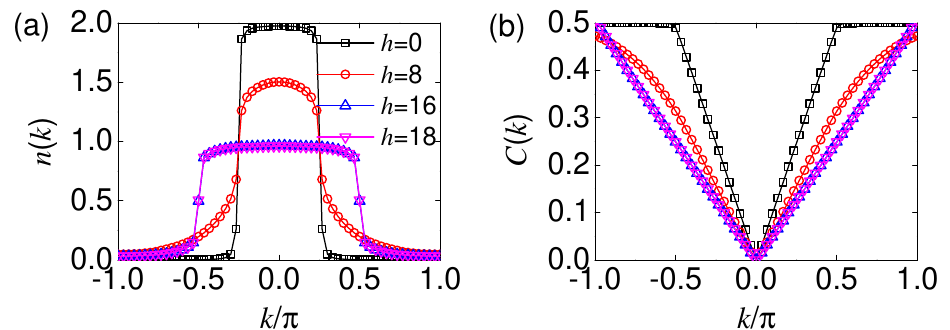}
\caption{(a) The momentum distribution function $n(k)$ and (b) the charge
structure factor $C(k)$ for systems with $\protect\rho =0.5$, $U=0$, $L=60$
and a varying $h$. }
\label{Fig9}
\end{figure}

\subsection{Momentum distribution of the conduction electrons}

The interplay of the localized and itinerant behaviors in the IKL can be
alternatively characterized by the momentum distribution function of the
conduction electrons,
\begin{equation}
n(k)=\frac{1}{L}\sum_{l,j}\left\langle \hat{c}_{l\sigma }^{\dag }\hat{c}_{j\sigma
}\right\rangle e^{i(l-j)k}
\end{equation}
Figure~\ref{Fig9}(a) shows the variation of $n(k)$ with increasing $h$ for $%
\rho =0.5$. For small $h$, as expected, the Fermi surfaces of the conduction
electrons is clearly formed at $k_{F}=\pm \pi \rho /2$. As $h$ increases,
the conduction electrons start to be hybridized with the local moments,
which broadens the distribution of $n(k)$ and blurs the original Fermi
surfaces. For even larger $h$, two sharp edges in $n(k)$, separating the
occupied and unoccupied states, appear at $k_{F}^{L}=2k_{F}=$ $\pm \pi \rho $. Further increasing $h$ has little effect on the structure of $n(k)$,
indicating that the system has reached a strong coupling regime with a
well-defined large Fermi surface. The variations in the structure of $n(k)$
is accompanied by corresponding changes in the charge structure factor,
defined by
\begin{equation}
C(k)=\frac{1}{L}\sum_{l,j}\left( \left\langle \hat{n}_{l}\hat{n}_{j}\right\rangle
-\left\langle \hat{n}_{l}\right\rangle \left\langle \hat{n}_{j}\right\rangle \right)
e^{i(l-j)k}.
\end{equation}
As shown in Fig.~\ref{Fig9}(b), we plot $C(k)$ for the same values of $h$ as
those used in Fig.~\ref{Fig9}(a). When $h$ is small, $C(k)$ exhibits two
cusps at $k_{c}=\pm \pi \rho $, manifesting the free particle nature of the
conduction electrons. The cusps, however, is smoothed at intermediate
values of $h$, exhibiting characters of strongly interacting fermions. Moving
into the strong coupling regime with larger $h$, two new cusps are formed at
$k_{c}^{L}=\pm 2\pi \rho $. Through a careful analysis of $n(k)$ and $C(k)$
for different filling number, we find that the appearances of the
characteristic wave vectors $k_{F}^{L}$ and $k_{c}^{L}$ are always
accompanied by the formation of FM. The above signatures of $n(k)$ and $C(k)$
strongly suggest that, in the strong coupling regime, the conduction
electrons behave like free spinless fermions.

\begin{figure}[tp]
\includegraphics[width=7.0cm]{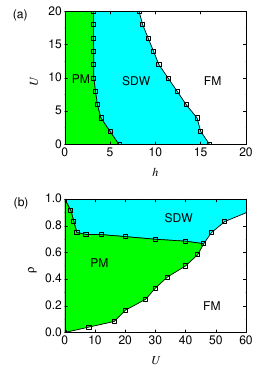}
\caption{(a) The phase diagram in the $h-U$ plane with $\protect\rho =0.75$
and (b) the phase diagram in the $\protect\rho -U$ plane with $h=4.0$. The phase boundaries in (a) and (b) have been extrapolated to the thermodynamic limit $L\rightarrow \infty $.}
\label{Fig10}
\end{figure}

To understand this, let us start from a half-filled IKL with $N$ $(=L)$
conduction electrons. In the strong coupling regime, each local moment tend
to catch an electron to form a local Kondo singlet, ending up with a 1D
liquid composed of $L$ singlets. Doping away from half filling amounts to
adding holes to the system. Note that double occupancy of holes on the same
lattice site is naturally excluded by definition. As discussed above, when
the spins of conduction electrons are aligned, the system can lower its
energy by allowing the holes freely moving on the lattice. Otherwise the
hoping process of holes may elevate energy due to the strong Kondo coupling.
Suppose now we have $L-N$ holes ($N<L$) moving in the sea of the singlet
background. The Fermi momentum of these mobile holes is thus $%
k_{F}^{h}=[1-(L-N_{c})/L]\pi =$ $\rho \pi =k_{F}^{L}$, exactly at the large
Fermi surface.

\begin{figure}[t]
\includegraphics[width=9cm]{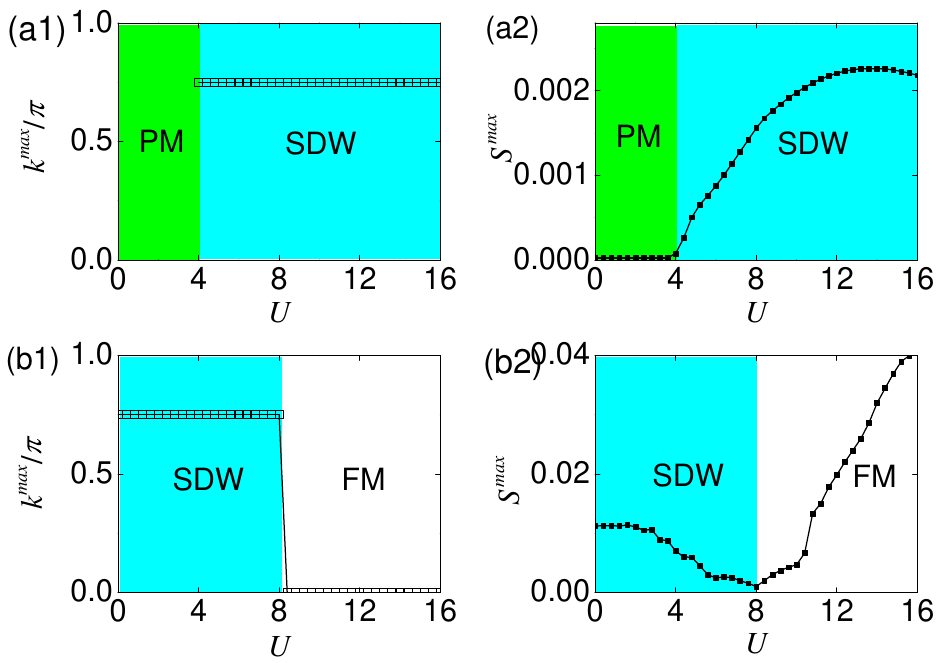}
\caption{(a1)-(b1) The ordering wave vector $k^{\max }$ and (a2)-(b2) the
ordering strength $S^{\max }$ as functions of $U$\ for systems with (a)\ $h=4.0
$ and (b) $h=12.0$.\ In these figures, we set $\protect\rho =0.75$.}
\label{Fig11}
\end{figure}

\section{Effect of the repulsive Hubbard interaction}
\label{sec:results2}

We proceed to discuss the effect of finite repulsive Hubbard interaction $U$%
. Following the same manner used in Sec.~\ref{sec:results1}, we present the
phase diagram in the $h-U$ plane for $\rho =0.75$ in Fig.~\ref{Fig10}(a). As
a direct impact of the Hubbard interaction, the FM becomes more robust, and
the phase boundary between FM\ and SDW is pushed to lower value of $h$. The
stabilization of FM by $U$ is also predicted in the isotropic KLM \cite%
{KLM7,ExactKLM3} and can be understood on physical grounds. The repulsive
interaction enhances the energy penalty costed by double occupancy of the
conduction electrons on the same lattice site, which promotes the formation
of the on-site Kondo singlet. The latter becomes mobile if the lattice is
doped far away from half filling, and thus favors ferromagnetism through the
Kondo mechanism we mentioned at the last of Sec.~\ref{sec:results1}. A more
interesting finding which does not exist in the isotropic KLM is that $U$
can also lower the critical value of $h$ between PM and SDW. This effect,
however, may saturate as $U$ increases to larger value.

To see clearly the role of $U$ in inducing various phase transitions, we
plot $k^{\max }$ and $S^{\max }$ as functions of $U$ for two different
values of $h$ in Fig.~\ref{Fig11}. It is to be seen that, depending on the
value of $h$, the repulsive interaction can drive both PM-to-SDW [Figs.~\ref%
{Fig11}(a1) and \ref{Fig11}(a2)] and SDW-to-FM [Figs.~\ref{Fig11}(b1) and %
\ref{Fig11}(b2)] transitions. We find again that, the ordering wave
vector $k^{\max }$ in the SDW is independent of $U$ and uniquely specified
by the Fermi momentum. As such, the phase transition driven by $U$ is of
course filling dependent. The general result is summarized in the $\rho -U$
phase diagram for $h=4.0$ [see Fig.~\ref{Fig10}(b)]. We observe that the
filling number influences the phase transitions in a similar fashion as
that in the $h-\rho $ phase diagram (see Fig.~\ref{Fig6}); the transition
from PM to FM turns out to be easier for low electron filling, and the SDW
is stabilized only closed to half filling.

\begin{figure}[t]
\includegraphics[width=9cm]{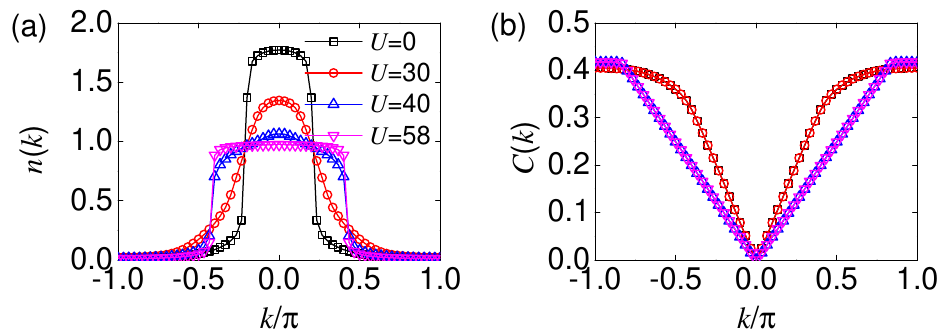}
\caption{(a) The momentum distribution function $n(k)$ and (b) the charege
structure factor $C(k)$ for systems with $\protect\rho =0.42$, $h=4.0$, $L=60$%
, and a varying $U$. }
\label{Fig12}
\end{figure}

Along the logic in Sec.~\ref{sec:results1}, let us examine the impact of $U$
on the momentum distribution function and charge structure factor. The
variations of $n(k)$ and $C(k)$ with $\rho =0.5$ and $h=4.0$ and a varying
values of $U$ are respectively shown in Figs.~\ref{Fig12}(a) and \ref{Fig12}%
(b). As depicted in these figures, when increasing $U$, both $n(k)$ and $%
C(k) $ exhibit the same behaviors as those shown in Fig.~\ref{Fig9}.
Especially when $U$ is sufficiently large for which the FM is reached, the
aforementioned large Fermi surface is clearly formed at $k_{F}^{L}=$ $\pm
\pi \rho $, suggesting a ferromagnetic metal. This is interesting since the
Hubbard interaction is commonly believed to render systems more insulating,
whereas its role here is reversed through the
aforementioned Kondo mechanism.

As can be found from above discussion, the role of repulsive Hubbard
interaction presents great similarities with the Kondo coupling $h$, hinting
that it may work through renormalizing $h$. The fully clarification of the
connection between $U$ and $h$ in the IKL may require some more
sophisticated analytical approaches, such as renormalization group and
bosonization, which is out of the scope of the present work and merits a
separate study in the future.

\section{Conclusion}

In conclusion, we have studied the ground-state properties of the 1D IKL by
using the numerical density-matrix-renormalizationgroup (DMRG) method. Three
distinct quantum phases, including a metallic PM, a matellic FM, and a
gapped SDW, have been obtained. The SDW is characterized by an ordering wave
vector which coincides with the nesting wave vector of the Fermi surface.
This makes the PM-to-SDW transition a magnetic analog of the Peierls
transition which occurs in the charge degree of freedom of a one-dimensional
metal. Moreover, by analyzing the momentum distribution function and charge
correlation function, the conduction electrons are shown to behave like
free spinless fermions in the FM. The effect of the
repulsive Hubbard interaction between conduction electrons has also been
clarified.

\section*{Acknowledgments}

This work is supported by the National Key R\&D Program of China under Grant
No. 2022YFA1404003, the National Natural Science Foundation of China (NSFC)
under Grant No.~12004230, 12174233 and 12034012, the Research Project
Supported by Shanxi Scholarship Council of China and Shanxi '1331KSC'.

\end{document}